# Theoretical and Experimental Investigations of High-Performance $Sr_2CoNbO_{6-\delta}$ Double Perovskite for IT-SOFC Cathode Applications

Jyotsana Kala[a,b], Vicky Dhongde[a,c], Suddhasatwa Basu[c], Brajesh Kumar Mani*[b], M. Ali Haider*[a,d]

[a] Renewable Energy and Chemicals Lab, Department of Chemical Engineering, Indian Institute of Technology Delhi, Hauz Khas, New Delhi-110016, Delhi, India

[b] Computational Many Body Physics Lab, Department of Physics, Indian Institute of Technology Delhi, Hauz Khas, New Delhi-110016, Delhi, India

[c] Fuel Cell Lab, Department of Chemical Engineering, Indian Institute of Technology Delhi, Hauz Khas, New Delhi-110016, Delhi, India

[d] Indian Institute of Technology (IIT) Delhi - Abu Dhabi, Khalifa City B, Abu Dhabi, United Arab Emirates

**Email:** bkmani@physics.iitd.ac.in, haider@iitd.ac.in

Enhancing the transport of oxygen anions in the cathode while maintaining surface stability is crucial for improving the performance of intermediate-temperature solid oxide fuel cells (IT-SOFCs). In this study, we investigated a novel cathode material, $Sr_2CoNbO_{6-\delta}$ (SCNO), using density functional theory (DFT), molecular dynamics (MD), and experimental characterization. Surface energy calculations using DFT indicated that the Nb cation at the B′-site suppresses Sr-cation segregation towards the surface but was also observed to inhibit the oxygen vacancy formation. However, redox-active Co at the B-site and Nb at the B′-site in SCNO together enhances surface stability and electrocatalytic performance. In addition, the SCNO structure was observed to be richer with oxygen vacancies at the surface than the bulk. Surface energies analysis for slabs with an oxygen vacancy predicts the migration of oxygen vacancies towards the surface, suggesting higher oxygen anion diffusivity for the surface than bulk. This prediction is supported by the MD simulation, which shows a higher oxygen diffusion rate for the surface than in the bulk. The experimental condition, which includes compressive strain on SCNO (deposited onto a 20% gadolinium doped ceria (GDC) substrate), was simulated using MD. Compressively strain SCNO film showed further reduction in the Sr-cation segregation. These findings provide new insights into controlling Sr segregation in SCNO and contribute to a better understanding of the enhanced oxygen reduction reaction (ORR) activity and high surface stability of SCNO. Subsequently, SCNO was synthesized to examine its potential as a cathode material in SOFCs. To investigate its performance, we developed symmetric cells with uniform dense thin films of varying thicknesses (40 and 80 nm) using the pulsed laser deposition (PLD) technique. Electrochemical impedance spectroscopy (EIS) and distributed Relaxation time (DRT) analysis suggest that the bulk oxygen ion diffusion is a limiting factor for ORR in SCNO. The polarization resistance ($R_p$) for 40 and 80 nm dense thin film symmetric cells are observed between 0.329 - 0.241 Ω cm$^2$ and 1.095 - 0.438 Ω cm$^2$, respectively, in the temperature range of 773 - 973 K in air atmosphere. Full cell configuration of NiO-GDC|GDC|SCNO is observed to demonstrate a significantly high peak power density of 0.633 W/cm² at 973 K. Theory-guided study and design of SCNO perovskite oxide suggest it as a potential candidate for IT-SOFC cathode materials.

## 1 Introduction

Solid oxide fuel cells are highly efficient and versatile energy conversion devices that convert chemical energy into electricity[1]. SOFCs operate at high temperatures (1073-1273 K). The cathode material in SOFCs plays a vital role in ORR for oxygen incorporation and transport[2,3]. The high operating temperature condition supports good ORR activity but necessitates the use of stable and costly materials for SOFC cathodes. Lowering the operating temperature results in sluggish ORR activity, limiting the overall performance of SOFCs. To overcome this challenge and improve the efficiency of SOFCs, developing novel cathode materials with enhanced ORR activity and oxygen ion diffusion at the intermediate temperature range (< 1073 K) is essential. As a result, significant research efforts have been devoted to exploring novel IT-SOFC cathode materials to reduce their cost and increase stability[4,5]. In recent years, extensive research has focused on double perovskite structures, specifically $A_2BB'O_{6-\delta}$ (B/B'= 3d, 4d, and 5d transition metals), to address the need for highly efficient cathodes that can transport oxygen ions effectively[6,7]. $Sr_2BB'O_{6-\delta}$, type of double perovskite oxide materials have been widely studied for electronic and magnetic properties for various applications[8–11]. The high structural stability and magnetic ordering of $Sr_2BB'O_{6-\delta}$ are leading research interests to probe their applicability to various applications. Researchers have reported $Sr_2BB'O_{6-\delta}$-type double perovskites as potential candidates for electrode applications in SOFCs. Some double perovskite oxides developed as electrode materials for SOFCs are $Sr_2MgMoO_{6-\delta}$[12], $Sr_2Fe_2O_{6-\delta}$[13,14], and $Sr_2Fe_{1.5}Mo_{0.5}O_{6-\delta}$[15]. Tian et al. reported that co-doped $Sr_2FeNbO_{6-\delta}$ is a chemically compatible cathode material with high conductivity and low polarization resistance for IT-SOFCs[16]. $Sr_2MgMoO_6$ and $Sr_2MnMoO_6$ are reported to show high-performance power density of 838 mW/cm$^2$ and 658 mW/cm$^2$ at 1073 K, respectively, for anode applications[12]. $Sr_2CoNbO_{6-\delta}$ is reported to show non-stoichiometry of oxygen, which is crucial for cathodic applications[17–19]. In addition, $Nb^{5+}$ doping at B-site of $SrCoO_{3-\delta}$ is also reported to show a lower area-specific resistance (ASR) of 0.16 and 0.68 $\Omega/cm^2$ at 773 and 723 K, respectively[20].

Structures incorporating strontium (Sr) at the A-site, exhibit good electrical conductivity, enhancement of oxygen defects, and performance of fuel cell[21,22]. However, it has been noted that the Sr at A-site tends to segregate from the bulk and accumulate on the surface, forming insulating phases (SrO)[23]. This can have a detrimental effect on the electrochemical performance as it tends to reduce the oxygen incorporation and oxygen diffusion on the surface by blocking the way for incoming oxygen molecules[23]. For example, Sr cation segregation in perovskites cathodes, such as $La_{1-x}Sr_xMnO_{3-\delta}$ (LSM)[24,25], $La_{1-x}Sr_xCo_{1-y}Fe_yO_{3-\delta}$ (LSCF)[26], $La_{1-x}Sr_xCoO_{3-\delta}$ (LSC)[27] and $LnBa_{1-x}Sr_xCo_2O_{6-\delta}$[28,29], is observed to affect the long-term stability and electrochemical performance of the cathode. Previous studies have shown that the extent of Sr segregation is affected by various factors, such as materials' oxygen non-stoichiometry[30], composition[25,28], temperature[25], operating condition[31], and lattice strain[28,32]. For instance, Koo *et al.* have proposed various strategies to control Sr surface segregation, such as imparting strain by lowering pO$_2$, iso-valent substitution, non-stoichiometric compositions, and surface modification[33]. In addition, Tsvetkov *et al.*[34] and Ramírez *et al.*[9] explored doping with high oxidation state and less reducible cations, like $Al^{3+}$, $Nb^{5+}$, $Hf^{4+}$, $Ti^{4+}$, $Sb^{5+}$, $Mo^{6+}$, or $Zr^{4+}$, to stabilize the perovskite structure, improve oxygen surface exchange kinetics, and prevent Sr surface segregation. Previous studies performed on $La_{0.5}Sr_{0.5}CoO_{3-\delta}$ had also demonstrated that Nb doping enhances the oxygen ion diffusion coefficient[35]. These findings suggest that $Sr_2CoNbO_{6-\delta}$ as a material of interest for IT-SOFC cathodes. A deeper understanding of the oxygen reduction reaction (ORR) is necessary to further improve these materials. This knowledge could enhance electrocatalytic activity and reduce Sr cation segregation in SCNO.

Alongside composition, fabrication techniques significantly influence the performance of cathode materials. Conventional fabrication techniques like tape casting, screen printing, and dip coating often face challenges. These include difficulties in controlling film thickness, uniformity, and porosity[36]. In addition, the films prepared with these techniques result in low oxygen vacancy concentrations and challenges in creating thin, multilayer films[37]. This affects the performance of the fuel cells. Pulsed laser deposition (PLD) offers a promising alternative. It allows precise control over microstructure and composition while maintaining high uniformity and optimizing the triple-phase boundary[38]. PLD, on the other hand, enhances ionic conductivity by forming heterogeneous structures and improving SOFC performance[39]. For example, Choi et al. demonstrated this by using PLD to prepare $PrBa_{0.5}Sr_{0.5}Co_{1.5}Fe_{0.5}O_{5+\delta}$ thin films (~100 nm thick) on a $BaZr_{0.4}Ce_{0.4}Y_{0.1}Yb_{0.1}O_3$ electrolyte. The PLD-produced films achieved a peak power density of ~1098 mW/cm², significantly higher than the ~800 mW/cm² achieved by films made with conventional methods at 873 K[40]. This also reduced offset resistance by improving cathode-electrolyte contact[40]. Despite its advantages, only a few studies have focused on depositing thin films of double perovskite oxides using PLD for SOFCs.

In the present work, we have investigated SCNO, a double perovskite material, as a candidate for IT-SOFC cathode applications. We started with a theoretical study of electrocatalytic properties and surface stability. DFT and MD simulations showed that SCNO exhibits high surface stability, oxygen vacancy concentrations, and oxygen anion diffusion. Subsequently, SCNO was synthesized experimentally, and a comprehensive characterization was performed, including the analysis of phase composition, surface morphology, oxidation states, and electrochemical performance as a cathode in SOFCs. To investigate rate-limiting factors in the ORR activity, symmetric cells were fabricated with 40 nm and 80 nm thicknesses using PLD. The rate-limiting processes in these thin-film symmetric cells were examined using EIS and DRT analysis. By combining theoretical insights with experimental design, we explored the underlying factors contributing to the high surface stability and enhanced performance of SCNO and found it as a potential candidate for IT-SOFC cathodes.

## 2 Methodology

### 2.1 Theoretical simulations

Spin-polarized density functional theory simulations were performed using the plane-wave pseudopotential approach implemented in the Vienna Ab Initio Simulation Package (VASP-5.4.4)[41,42]. We used the projector augmented wave method (PAW) to account core electronic states and the Perdew-Burke-Ernzerof functional (GGA-PBE) as the exchange-correlation potential[43,44]. All simulations were performed with an energy cut-off of 520 eV. Hubbard correction was included to predict the accurate behavior of strongly localized d-electrons in Co and Nb. The U values were obtained using the Cococcioni method[45]. The calculated U values for Co and Nb are 5.45 and 4.01 eV, respectively (Figure S1). The convergence criteria of $10^{-6}$ eV and 0.05 eV/Å were used for electronic and ionic relaxation, respectively. We used $3 \times 3 \times 3$ and $3 \times 3 \times 1$ gamma-centered k-mesh grids in computations for bulk and slabs, respectively.

To trace the dynamics of ions, molecular dynamics simulations were performed on $Sr_2CoNbO_{5.75}$ using the Large-scale Atomic/Molecular Massively Parallel Simulator (LAMMPS, Sandia National Laboratories, USA)[46]. Buckingham inter-atomic potentials[47] were used to compute the interactions of cations with oxygen anion, with a cut-off radius of 20 Å. The coulombic and van der Waals interactions in the potential are expressed as

$$\varphi(r_{ij}) = \frac{z_i z_j}{4\pi\varepsilon_0} \frac{e^2}{r_{ij}} + A_{ij} e^{-r_{ij}/\rho_{ij}} - \frac{C_{ij}}{r_{ij}^6},$$

where the parameters A, ρ, and C represent the strength of the Buckingham interaction. The parameters z and $r_{ij}$ represent the ionic charge and interionic separation, respectively. The coulombic interactions were computed using particle-particle particle mesh with $10^{-4}$ eV/Å relative error in forces[48,49]. The interaction parameters used in our MD simulations are provided in Table S1. We used the Nose-Hoover thermostat[50] and Verlet algorithm[51] to keep the temperature constant and integrate the equation of motion, respectively. To remove the finite size effects, we have used a larger supercell of $8 \times 8 \times 8$ with 9,984 atoms. For surface properties, the slabs were terminated along the (001) direction with a 100 Å vacuum. For both bulk and surface simulations, the structure was equilibrated in NPT ensemble for 100 ps, then in NVT ensemble for 1 ns in the steps of 1 fs. The analysis was performed on 10 different MD runs using the same ensemble for 1 ns. The average of 10 production runs was considered while extracting the properties. Visual molecular dynamics program was employed to visualize the density profiles[49].

## 2.2 Experimental details

SCNO double perovskite was prepared using the combustion technique (citric acid- metal). A stoichiometric amount of $Co(NO_3)_2 \cdot 6H_2O$ (Alfa Aesar), $Sr(NO_3)_2$ (Alfa Aesar) and $C_4H_4NNbO_9 \cdot xH_2O$ (Sigma Aldrich) was dissolved in 100 ml of deionized water. Then citric acid (Alfa Aesar) and polyethylene glycol (PEG-400, Alfa Aesar) were added to the above solution as a complexing agent under continuous stirring. The molar ratio of metal nitrate: PEG-400: citric acid was maintained as 1:1:1.5. The complex solution was stirred for 12 h at 363 K to evaporate water and gel formation. Further, the formed gel was self-combusted at 573 K to obtain a preliminary powder, followed by sintering at 1573 K for 48 h in the air to achieve the desired phase. The SCNO target was prepared by adding synthesized double perovskite in a one-inch-diameter dye set and pelletizing the powder using a hydraulic press at 200 kg/cm$^2$ pressure. Then, the obtained pellet was calcined at 1473 K for 6 h and polished using an emery paper to establish a close connection with the target holder. The gadolinium doped ceria (GDC, Fuel cell materials, USA) powder was mixed with a binder of 15% polyvinyl alcohol solution, followed by drying at 353 K overnight. The dry GDC was ground into a fine powder utilizing an automatic mortar pestle and pelletized in a dye set using a hydraulic press. The GDC pellets were heated at 1723 K for the duration of 10 h in a furnace under an air atmosphere and used further as an electrolyte to fabricate symmetric cells.

On GDC pellets, symmetric cells with thin film electrodes were developed using the PLD technique. We have optimized growth parameters, such as deposition temperature, target-to-substrate distance, post-deposition annealing temperature, oxygen partial pressure, oxygen pressure, laser fluence, and time, to accomplish the highest film quality. The thin film was formed under the following optimal conditions: Target-to-substrate distance was 5 cm, oxygen partial pressure was $10^{-3}$ mbar, and base pressure was maintained at $9 \times 10^{-10}$ mbar. On the opposite side of the electrolyte, a similar deposition was made. After depositing, the pellet was annealed at 1473 K for an hour to acquire the stable electrode-electrolyte interface. Apart from this, SCNO was mixed with 10% PVA solution and dried in the oven at 353 K, followed by grinding into fine powder. For slurry preparation, the obtained SCNO powder was mixed with α-terpinol and ethyl cellulose, then pasted on a GDC pellet using a brush painting technique followed by calcination for 6 h at 1473 K.

An anode-supported single cell (NiO-GDC|GDC|SCNO-GDC) was fabricated by solid-state palletization of mixing 40 wt% nickel oxide (NiO, Alfa aesar), 59 wt% GDC, and pore former 1 wt% poly methyl methacrylate (PMMA, Alfa aesar) and added to the dye set. The GDC powder was layered on the NiO-GDC mixer, then palletized and sintered to 1723 K for 10 h. Similarly to NiO-GDC, the SCNO-GDC (SCNO 69% - GDC 30% and 1% PMMA) was prepared and coated on the obtained pellet (NiO-GDC|GDC) using brush painting, and then the cell fired at 1473 K for 6 h.

The as-synthesized SCNO double perovskite structure was analyzed using X-ray diffraction (XRD, Rigaku, MiniFlex 600). The structural information of as-synthesized SCNO was evaluated using FullProf software[54]. The morphological and microstructural analysis of SCNO powder, thin-film, and brush-painted cells were analyzed using the field-emission scanning electron microscopy technique (FESEM, TESCAN MIRA3, Czech Republic). The valence oxidation states of elements involved in the formation of double perovskite and its surface compositions were examined using X-ray photoelectron spectroscopy (XPS, Axis Supra).

The symmetric cells of 40 and 80 nm, and brush-painted cells EIS were carried out in the air to study the rate-limiting step of the oxygen reduction reaction. However, the challenge was associated with distinguishing various electrode processes because of their overlapping impedance contributions. Therefore, MATLAB code was used to effectively differentiate the complex electrode reaction mechanism to investigate EIS data using the DRT method[55]. In addition, an anode-supported full-cell, i.e., NiO-GDC|GDC|SCNO performance, was estimated in a split furnace temperature range of 773 - 973 K. The ambient air was utilized as an oxidant at the cathode side. Humidified 3 vol% of $H_2O$ hydrogen was used as fuel on the anode side.

## 3 Results and Discussion

SCNO material was simulated using DFT simulations to probe electrocatalytic properties. The initial crystal structure information was taken from the prior experimental studies[19,56,57]. Using these experimental inputs, a unit cell of SCNO consisting of 20 atoms was constructed and subsequently optimized using DFT+U, to obtain a global minimum structure. Obtained lattice parameters, along with the available data from the literature for comparison, are provided in Table 1. As evident from the table, the computed lattice parameters are in good agreement with the literature values.

**Table 1** Optimized lattice parameters for SCNO, SCO, and SNO from DFT simulations and literature.

| Material | a (Å) | b (Å) | c (Å) | Ref. |
| --- | --- | --- | --- | --- |
| $Sr_2CoNbO_6$ | 5.56 | 5.56 | 7.87 | This Work |
|  | 5.58 | 5.58 | 7.93 | [Ref.56] |
|  | 5.58 | 5.58 | 7.91 | [Ref.19] |
|  | 5.58 | 5.58 | 7.87 | [Ref.57] |
| $Sr_2Co_2O_6$ | 5.56 | 5.56 | 7.56 | This Work |
| $Sr_2Nb_2O_6$ | 5.75 | 5.75 | 8.20 | This Work |

The electrocatalytic properties of bulk SCNO are studied using DFT calculations with Hubbard correction to incorporate the effect of strongly correlated d-electrons of Co and Nb. Since oxygen vacancies play a major role in oxygen incorporation and transport, the oxygen vacancy formation energy ($E_{OV}$) could acts as a descriptor for the electrocatalytic performance of the material. Considering this, $E_{OV}$ in possible oxygen vacancy sites is computed and analyzed in the SrO and CoNbO planes (Figure 1(a)). The $E_{OV}$ was obtained using the expression

$$E_{OV} = E_{defect} + 0.5 \times E_{O_2} - E_{perfect},$$

where $E_{perfect}$, $E_{defect}$, and $E_{O_2}$ represent DFT energies associated with stoichiometric, non-stoichiometric structures, and gas-phase oxygen molecule, respectively.

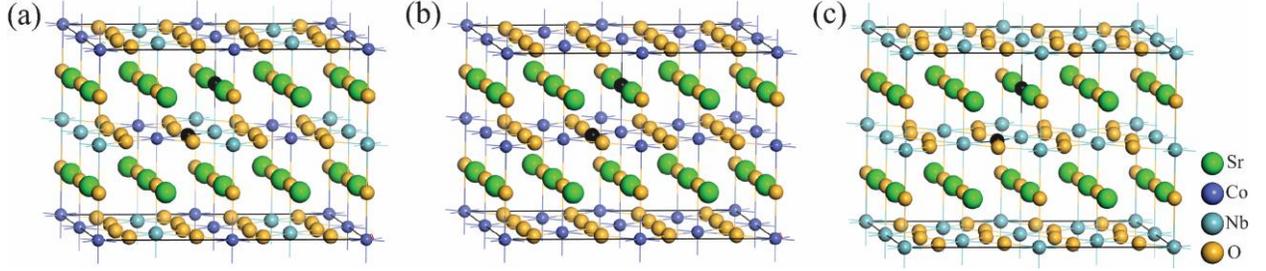

**Figure 1.** 2 × 2 × 1 supercells of (a) SCNO, (b) SCO, and (c) SNO structures with the oxygen vacancy site depicted as the black sphere in the (Co and/or Nb)O and SrO planes.

**Table 2** The oxygen vacancy formation energies ($E_{OV}$) for SCNO, SCO, and SNO structures in different crystal planes.

| Material | Oxygen Vacancy Plane | $E_{OV}$ (kJ/mol) |
|---|---|---|
| $Sr_2CoNbO_{6-\delta}$ | CoNbO | 158.73 |
| | SrO | 107.53 |
| $Sr_2Co_2O_{6-\delta}$ | CoO | -41.74 |
| | SrO | -37.33 |
| $Sr_2Nb_2O_{6-\delta}$ | NbO | 220.05 |
| | SrO | 194.69 |

Calculated values of $E_{OV}$s' are provided in Table 2. The $E_{OV}$ in SrO and CoNbO planes are estimated to be 158.73 kJ/mol and 107.53 kJ/mol, respectively. This suggests oxygen vacancies formation to be more facilitated in the SrO plane than in the CoNbO plane. We anticipate the presence of Nb cations in the B′O plane to be responsible for the decreased favorability of oxygen vacancy creation. In order to investigate the effect of Nb on oxygen vacancy formation, system was remodeled with only Co or Nb cation at the B and B′-site, resulting in the formation of $Sr_2Co_2O_6$ (SCO) and $Sr_2Nb_2O_6$ (SNO), respectively. SCO and SNO structures were subsequently fully optimized using DFT+U calculations. The lattice parameters for these systems are given in Table 1. The CoO plane of SCO (-41.74 kJ/mol for CoO and -37.33 for SrO plane) is observed to be the most favorable for oxygen vacancies, Figure 1(b). The study performed by Wexler *et al.* reported the impact of various transition metals at the B-site on $E_{OV}$[58]. The group also reported negative $E_{OV}$'s in SrCoO$_3$[58]. On the other hand, $E_{OV}$ in NbO plane (220.05 kJ/mol for NbO and 194.69 for the SrO plane) of SNO is observed to be the highest among all the structures, Figure 1(c). The higher redox ability of Co than Nb could be a possible reason for the observed trend.

In order to probe the redox ability of ions in SCNO as the possible reason for the observed trend, the Bader charge analysis is performed on the perfect and oxygen-vacant structure of SCNO. Table 3 encapsulates the average Bader charges associated with Sr, Co, Nb, and O ions in SCNO. A change in charges associated with surrounding ions after the formation of oxygen vacancy due to the redistribution of electron densities. However, the charge associated with Sr (+1.59) remain the same with or without oxygen vacancy formation. This could be correlated with the redox inactivity of Sr cation[59]. Along with Sr, Nb ions present in the BO plane had also been observed to be redox inactive. The high oxidation state of Nb$^{5+}$ and low electronegativity, along with redox inactivity, could be responsible for a high work function for oxygen vacancy creation[60,61]. The studies performed by Greiner *et al.* [61], Wang *et al.* [62], and Hu *et al.* [63] also reported Nb cations resulting in higher oxygen vacancy formation energies. On oxygen vacancy formation in the SrO plane or CoNbO plane, Co ions are observed to reduce more, hence favoring the oxygen vacancy creation in the structure[58]. As also evident from the Bader charges, Nb cations are reduced by only 1.42 and 1.78% while Co cations are reduced by 3.57 and 4.28% on oxygen vacancy creation in CoNbO and SrO planes, respectively. On the other hand, the redox ability of Co ions (4.44% and 5.66% on oxygen vacancy creation in the CoO and SrO plane, respectively) in SCO favors oxygen vacancies by reducing the energy penalty for oxygen vacancy creation, and the redox inability of Nb$^{5+}$ (only 2.82%) hinders the formation of oxygen vacancies in SNO.

**Table 3** Bader charges associated with Sr, Co, Nb, and O ions in SCNO, SCO, and SNO systems.

| Material | Oxygen vacant plane/Ion | Sr | Co | Nb | O |
| --- | --- | --- | --- | --- | --- |
| Sr$_2$CoNbO$_{6-\delta}$ | Perfect | 1.59 | 1.40 | 2.81 | -1.23 |
| | CoNbO | 1.59 | 1.35 (3.57%) | 2.77 (1.42%) | -1.24 |
| | SrO | 1.59 | 1.34 (4.28%) | 2.76 (1.78%) | -1.24 |
| | Perfect | 1.60 | 1.59 | - | -1.07 |
| $Sr_2Co_2O_{6-\delta}$ | CoO | 1.60 | 1.52 (4.44%) | - | -1.06 |
| | SrO | 1.60 | 1.50 (5.66%) | - | -1.06 |
| $Sr_2Nb_2O_{6-\delta}$ | Perfect | 1.59 | - | 2.48 | -1.36 |
| | NbO | 1.58 | - | 2.41 (2.82%) | -1.36 |
| | SrO | 1.58 | - | 2.41 (2.82 %) | -1.36 |

To get further insights into the favorable pattern of oxygen vacancy sites, increased concentration of oxygen vacancies is considered for estimation of $E_{OV}$. In order to find the minimum energy non-stoichiometric structures with increasing concentration of oxygen vacancies, $E_{OV}$ for possible oxygen vacant sites in the SrO and CoNbO planes are calculated, Figure 2(a). The data on $E_{OV}$s' with concentrations is given in Table S2. It was observed that with the increasing concentrations of oxygen vacancies, the $E_{OV}$ shows the same trend among CoNbO and SrO planes as observed for the case of $\delta = 0.125$. The $E_{OV}$s' are observed to be 158.73 kJ/mol, 276.21 kJ/mol, and 491.39 kJ/mol for the CoNbO plane and 107.53 kJ/mol, 161.28 kJ/mol and 354.53 kJ/mol for SrO plane for $\delta = 0.125$, 0.25, and 0.375, respectively. The linear trend of $E_{OV}$ with oxygen non-stoichiometry $\delta = 0.125$-$0.375$ suggests stability of the perovskite structure under these oxygen vacancy concentrations. The $E_{OV}$ is observed to be very high for $\delta = 0.5$ (656.89 kJ/mol for the CoNbO plane and 437.60 kJ/mol for the SrO plane). This implies that the oxygen vacancy content of $\delta = 0.5$ and beyond could only be achieved at the cost of a very high energy penalty. This could lead to the saturation of oxygen vacancy concentration in SCNO to $\delta = 0.375$ and instability of perovskite oxide after increasing the oxygen vacancy content. Figure 2(b, c) represents the most favorable oxygen vacancy sites for $\delta = 1$ in CoNbO and SrO planes, respectively. As discernible from the Figure, the vacancies prefer to form row-wise in available sites, making a tunnel-like pathway for oxygen anion transport in the material. Similar observations were reported by Maiti et al.[64] on perovskite oxides, and Prassana et al.[65] and Kim et al.[66] in brown millerite structures.

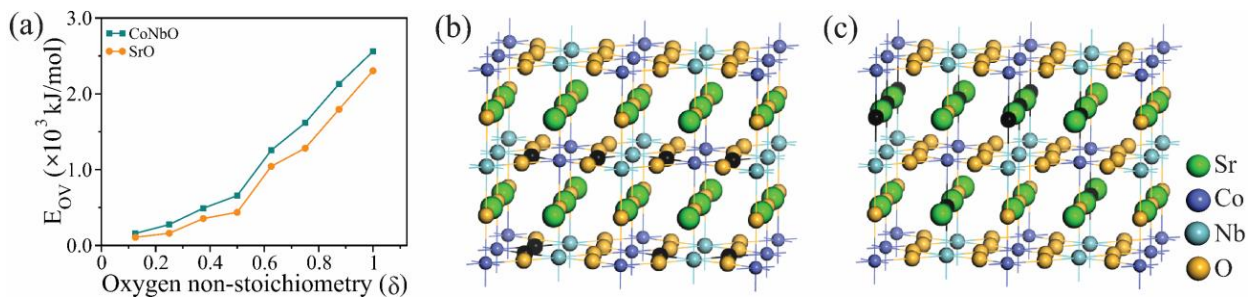

**Figure 2.** (a) Trend of $E_{OV}$ with increasing oxygen non-stoichiometry ($\delta = 0.125\text{-}1$). The favorable vacancies (depicted as black color sphere) for SCNO in (b) the CoNbO and (c) the SrO planes.

MD simulations are performed to estimate the oxygen anion diffusivity in bulk SCNO. Since oxygen vacancy formation energies are considered descriptor for oxygen anion diffusivity[67], SCNO structure is modeled with $\delta = 0.25$[18]. In order to maintain the overall charge neutrality, the Co ions oxidation state is modeled to be +2.5. The oxygen anion diffusivity is obtained using mean square displacement (MSD) vs. time data, as shown in Figure 3(a), using the expression

$$MSD = 6Dt + B,$$

where D, B, and t represent the oxygen anion diffusivity, a constant, and time, respectively.

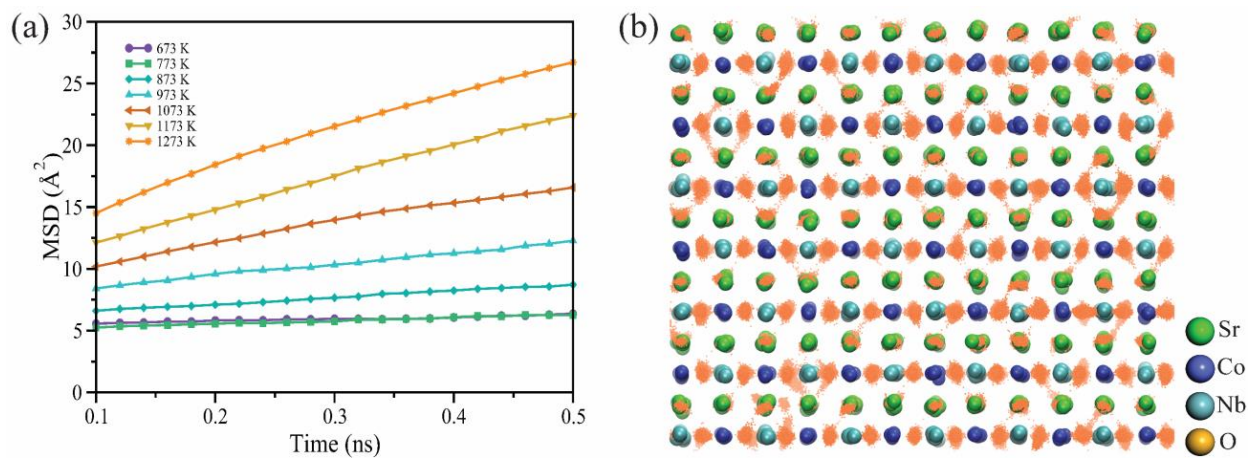

**Figure 3.** (a) MSD vs. time data extracted on equilibrated structures, (b) oxygen anion density profile of SCNO at 873 K temperature.

**Table 4** Oxygen anion diffusivity in SCNO extracted from MSD vs. time data for 1 ns on equilibrated structures.

| Temperature (K) | D ($\times 10^{-7}$ cm$^2$ sec$^{-1}$) |
|---|---|
| 673 | 0.22 ± 0.19 |
| 773 | 0.44 ± 0.24 |
| 873 | 0.86 ± 0.23 |
| 973 | 1.56 ± 0.31 |
| 1073 | 2.69 ± 0.22 |
| 1173 | 4.38 ± 0.12 |
| 1273 | 5.02 ± 0.27 |

As evident from Table 4, the oxygen anion diffusivity is thermally activated in SCNO. Despite the presence of $Nb^{5+}$, hindering the formation of oxygen vacancies, it was observed that oxygen anion diffusivity for SCNO ($8.6 \times 10^{-8}$ cm$^2$/sec at 873 K temperature) comparable with the other potential cathode materials like $La_{1-x}Sr_xCoO_{3-\delta}$ ($2.73 \times 10^{-9}$ cm$^2$/sec[35] for $La_{0.5}Sr_{0.5}CoO_{3-\delta}$ and $5.47 \times 10^{-9}$ cm$^2$/sec[35] for $La_{0.5}Sr_{0.5}Co_{0.95}Nb_{0.05}O_{3-\delta}$ at 873 K), $LnBaCo_2O_{6-\delta}$ ($5 \times 10^{-8}$ cm$^2$/sec[68] for $GdBaCo_2O_{5.5}$, $7.5 \times 10^{-8}$ cm$^2$/sec[68] for $GdBa_{0.5}Sr_{0.5}Co_{2-x}Fe_xO_{6-\delta}$, $1.76 \times 10^{-7}$ cm$^2$/sec[28] for $NdBaCo_2O_{6-\delta}$ at 873 K). Activation energy of oxygen anion diffusion in SCNO is estimated using the Arrhenius equation for thermally activated diffusivities

$$D = D_0 e^{-\frac{E_a}{k_B T}},$$

where $E_a$ and $k_B$ represent the activation energy associated with oxygen anion diffusion and Boltzmann constant, respectively. $D_0$ and T represent the pre-exponential factor and temperature, respectively. The activation energy for oxygen anion diffusion associated with SCNO is estimated as 38.48 kJ/mol (Figure S2). This is observed to be improved when compared with reported layered perovskite SOFC cathodes like $NdBaCo_2O_{6-\delta}$ (40.82 kJ/mol[28]), $GdBaCo_2O_{6-\delta}$ (50.8 kJ/mol[68]) and $PrBaCo_2O_{6-\delta}$ (98.4 kJ/mol[69]).

Surface $E_{OV}$ could provide a better description of the electrocatalytic performance of SCNO as oxygen anion incorporation reaction takes place on the surface. To calculate surface $E_{OV}$, SCNO surface slabs with CoNb/Sr and Sr/CoNb terminations (where A/B represent A ions present on the top-most layer and B ions present in the sublayer of the slab) were created along (001) direction (shown in Figure 4(a, b)). Subsequently, oxygen vacancies are created in the respective top surfaces of the slabs. A lower surface $E_{OV}$ on the surface (27.61 kJ/mol for CoNb/Sr and 49.74 kJ/mol for Sr/CoNb slabs) is observed as compared to the bulk (Table 2). A similar trend of a lower surface $E_{OV}$ is reported by Zhukovskii *et al.* in $PbTiO_3$, $PbZrO_3$, and $SrTiO_3$ perovskites oxides[70]. Previous studies performed by Walsh *et al.* have shown higher $E_{OV}$ for higher coordination numbers in $In_2O_3$[71]. For example, the $E_{OV}$ for a site with 4 coordination number is reported as 288.96 kJ/mol, whereas for the oxygen vacancy site with 3 coordination number is reported to be 207.36 kJ/mol[71]. The oxygen vacancy sites at the surface and bulk differ by coordination number. An opposite trend of oxygen vacancy formation energies is observed for CoNb/Sr and Sr/CoNb terminated surface slabs than bulk. The $E_{OV}$ for the top surface layer of the CoNb/Sr terminated slab (27.61 kJ/mol) is observed to be lower than that of the Sr/CoNb terminated slab (49.74 kJ/mol). It was therefore anticipated that the distorted bonding pattern of Nb − O on the surface was not able to influence the surface $E_{OV}$ as compared to the bulk in the CoNbO plane. Hence, the dominating nature of Co in terms of redox ability leads to the reduction in the surface $E_{OV}$s'.

The observed trend of $E_{OV}$ in bulk and surface, thus predicts higher oxygen vacancy concentration on the surface and, hence, probable oxygen vacancy defect migration towards the surface. This could further enhance surface oxygen anion diffusion by facilitating oxygen incorporation in SCNO. In order to study the surface oxygen vacancy formation in detail, surface energy calculations are performed with oxygen vacancy at the top surface, sublayer, and bulk ($3^{rd}$, $4^{th}$, $5^{th}$, and $6^{th}$) layer of CoNb/Sr surface terminated slab along (001) direction. The surface energies of the oxygen vacant slab are calculated using the expression

$$\gamma = \frac{1}{2A}(E_{surface} - n \times E_{bulk}),$$

where $E_{slab}$ represents the DFT energy of the CoNb/Sr terminated slab with oxygen vacancy at top layer, sublayer, and bulk layers. $E_{bulk}$, n, and A represent the energy of the bulk structure, the number of bulk cells in the slab, and the cross-sectional area of the slab. Surface energies provide a good description of the surface stability of a particular termination or orientation[72]. The corresponding structures with oxygen-vacant sites are shown in Figure 4(b). A comparative trend of surface energies for oxygen vacancies is shown in Figure 4(c), and values are provided in Table S3. The $E_{OV}$ is lowest on the top surface as compared to the sublayers and the bulk layers. This implies that the migration of oxygen vacancies towards the surface. As evident from Table S3, on going from top-surface to bulk, the favorability of oxygen vacancy is decreased. This predicts an enhanced performance on the thin film surface as compared to the bulk. The material is then expected to show higher oxygen surface exchange coefficient and surface oxygen anion diffusion. In order to investigate the oxygen anion diffusion on the surface, the CoNb/Sr and Sr/CoNb surface terminated slabs are modeled with 9,984 atoms and 100 Å vacuum in the z-direction. The time dynamics of oxygen anions are captured at every fs for 1 ns on equilibrated structures at 873 K. The corresponding surface oxygen anion diffusion is obtained as $2.91 \times 10^{-7}$ cm$^2$/sec and $2.79 \times 10^{-7}$ cm$^2$/sec for CoNb/Sr and Sr/CoNb terminated slabs, respectively, as compared to $1.39 \times 10^{-7}$ cm$^2$/sec for bulk structure at 873 K temperature (Figure S3). This is consistent with aforementioned DFT results on $E_{OV}$s' for bulk and surface slabs.

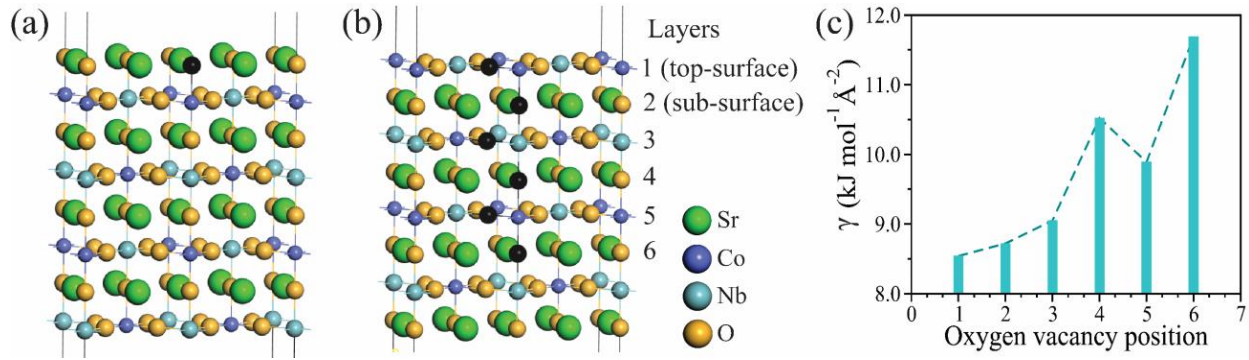

**Figure 4.** (a) Sr/CoNb surface terminated slab with oxygen vacancy on the top surface and (b) Co/Nb surface terminated slab with oxygen vacancy on the top surface, sub-surface, 3$^{rd}$, 4$^{th}$, 5$^{th}$, 6$^{th}$ layer in the bulk of SCNO slab, (c) Comparative trend of surface energies of oxygen vacant CoNb/Sr slab.

Sr-containing perovskites, as in the present case of SCNO, are reported to show degrading Sr-cation segregation tendencies. For example, $La_{1-x}Sr_xMnO_{3-\delta}$ (LSM)[24], $La_{1-x}Sr_xCo_{1-y}Fe_yO_{3-\delta}$

(LSCF)[26], and $LnBaCo_2O_{6-\delta}$[28,29] are reported to show degrading Sr-cation segregation at SOFC operating conditions, affecting the surface stability and hindering the use of these materials due to long-term stability issues. To probe Sr cation segregation, energetic favorability of possible terminations of the surface slab is examined. Figure S4 represents the Sr/CoNb and CoNb/Sr terminated slabs. Surface energy ($\gamma$) analysis of stoichiometric Sr/CoNb and CoNb/Sr terminating slabs is performed[28]. The surface energies of Sr/CoNb and CoNb/Sr are estimated to be 6.25 and 6.54 kJ/mol·Å$^2$, respectively. The surface energy of the Sr terminating slab Sr/CoNb is observed to be less than the CoNb/Sr surface slab. This implies tendency for Sr cation surface segregation in the material. In order to study the role of Nb ions in Sr-cation segregation, the model was extended from optimized SCNO to SCO and SNO slab structures, Figure S4. Sr/Co slab is observed to show relatively lower surface energy (4.18 kJ/mol·Å$^2$) as compared to the Co/Sr slab (4.65 kJ/mol·Å$^2$) in SCO, whereas Sr/Nb slab (5.44 kJ/mol·Å$^2$) has the comparable surface energy as compared to Nb/Sr (5.19 kJ/mol·Å$^2$) in SNO. It clearly indicates the role of Nb ions in enhancing surface stability. Hence, it is predicted that the optimum presence of Co and Nb cations at B and B′ sites help suppressing Sr cation segregation towards surface along with significantly enhanced oxygen vacancy formation.

In order to study the time dynamics of Sr-cation segregation in SCNO at 873 K, MD simulations are performed on a surface slab with CoNb/Sr termination. The surface slabs are modeled with a 100 Å vacuum in the z-direction. The ion dynamics in SCNO is traced at 873 K temperature for 1 ns. The corresponding ion density profile of the SCNO slab is shown in Figure 5 (a). For quantitative measurement, the degree of Sr cation segregation is estimated on the surface as the percentage of Sr-cation migrating from the sub-layer to the top-most layer[28]. The degree of cation segregation is reported as an important parameter for comparative segregation tendencies as well as quantitative measurement of cations separately in $NdBa_{1-x}(Sr/Ca)_xCo_2O_{6-\delta}$, [28]. In MD simulations, the degree of cation segregation is evaluated as the percentage of Sr cation moving from sub-surface to top surface in the slabs after 1 ns of MD run[28]. The SCNO slab consists of 128 Sr atoms in the sublayer for the CoNb/Sr-terminated layer. After ionic dynamics at a temperature of 873 K, an average of 28 Sr atoms migrated from the sublayer to the top surface. Consequently, the degree of Sr-cation segregation is estimated to be 21.875%. To account for the substrate effect, as in the experimental setup, SCNO films are remodeled in the presence of strain. The lattice mismatch strain between SCNO and gadolinium doped ceria (20%) GDC is calculated, as it is reported to be a compatible substrate electrolyte material for SOFCs[1]. Using the lattice parameters of GDC (a = 5.42 Å[73]) and SCNO (Table 1), around 2.5% compressive strain is computed. Compressively strained CoNb/Sr and Sr/CoNb terminated slabs are observed to have similar surface energies of 5.83 kJ/mol·Å$^2$ and 5.44 kJ/mol·Å$^2$, respectively. This predicts suppression of Sr-cation segregation in the SCNO film deposited on GDC substrate. In line with the DFT results, the degree of cation segregation as evaluated from MD. With the effect of compressive strain from the GDC substrate, only 8 Sr atoms migrated from the sublayer to the top surface at a temperature of 873 K. Consequently, the degree of Sr-cation segregation was observed to decrease to 6.25%. The corresponding ion density profile of SCNO with 2.5% lattice strain is shown in Figure 7 (b). SCNO film deposited on GDC are observed to be more stable with very less Sr-cation segregation. Higher oxidation number of Nb cations and its ability to promote the bonded pattern leads to a better surface stability in the structure. In addition, the compressed state of the film leads to less space available for the Sr cation to segregate on the surface of the film[28]. Theoretical investigation predicts SCNO films deposited on GDC to show significantly reduced Sr-cation segregation with improved electrocatalytic performance for IT-SOFCs.

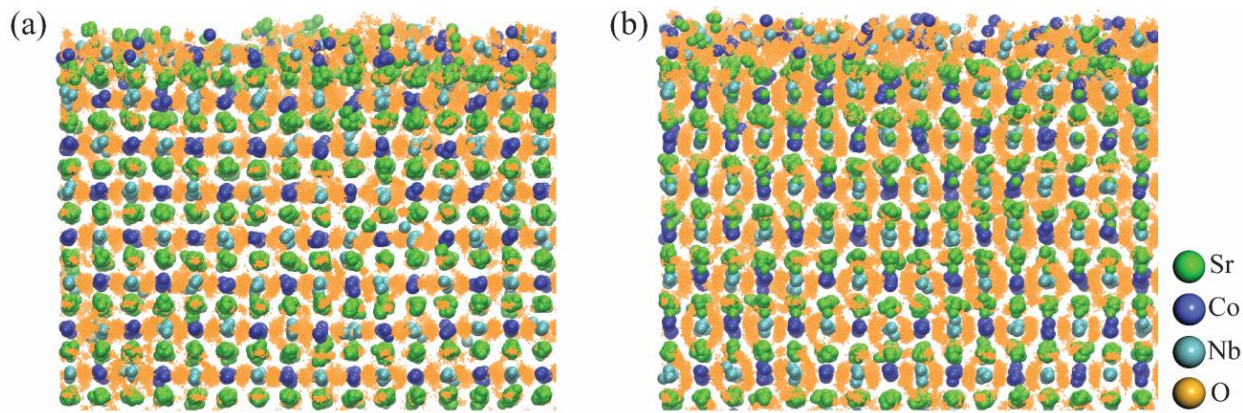

**Figure 5.** Cation distortion profiles for (Co, Nb)/Sr terminated slabs of SCNO in (a) unstrained and (b) compressively strained state at 873 K.

Subsequently, thin films of SCNO on GDC pallet are fabricated experimentally using PLD. The XRD patterns of 40 and 80 nm SCNO films are shown in Figure 6 (a). As discernible from the Figure, the GDC substrate shows distinct peaks that match well with its corresponding JCPDS: 75-016. The SCNO peak intensity is lower for the thin film than the bulk structure. The most prominent peak in the SCNO film appears at 31.91°, and the bulk structure shows the desired phase. The FESEM images in Figure 6 (b) revealed a dense surface with connected grain boundaries for the bulk SCNO. The cross-section images in Figure 6 (c, d) show a uniformly deposited dense film with a clear interface between the electrode and electrolyte, and thicknesses of approximately ~40 and ~80 nm, respectively. This indicates the formation of a geometrically well-defined structure. In order to probe the oxygen non-stoichiometry ($\delta$) in the structure, thermogravimetric analysis was performed. The calculated $\delta$ values at different temperatures and associated plot is shown in Table S4 and Figure S6. The $\delta$ values were observed in the range of 0.26-0.33 for temperature range 473-1073 K temperatures. This observation is in line with the $E_{OV}$ measurements from DFT (Figure 2, Table S2).

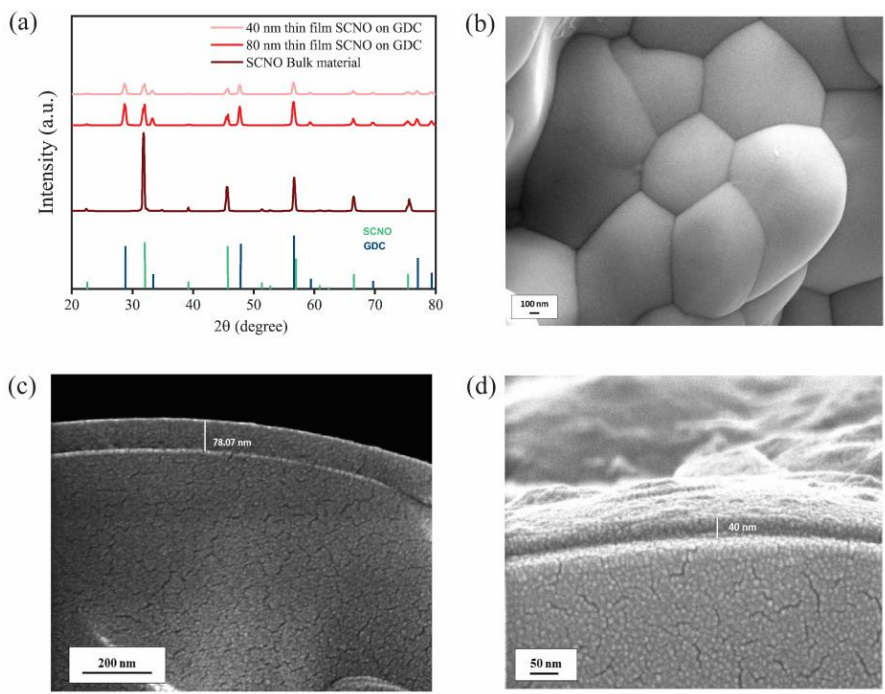

**Figure 6.** (a) XRD pattern of SCNO, 80 and 40 nm thin films of SCNO deposited over the GDC electrolyte using PLD, (b) FESEM image of the surface morphology of the SCNO bulk material, the cross-section of a thin film of SCNO on GDC substrate with (c) ~80 nm and (d) ~40 nm thickness.

The XPS spectrum was analyzed using SCNO as-synthesis and thin films to investigate the possible valence state charges of the associated elements. In addition, XPS spectrum is used to probe the surface [29,74] for Sr-cation segregation tendencies. Figure 7 (a) presents the spectra of Sr $3d$ electrons for bulk SCNO and thin films placed on a GDC electrolyte. The Sr spectra display two distinct peaks that align with the crystal lattice of the double perovskite structure. The peaks are characterized by energy values of 131.4 and 133.2 eV and are observed to be associated with $Sr^{2+}$ $3d_{5/2}$ and $3d_{3/2}$ lattices, respectively[75]. No evidence of Sr segregation was observed in the XPS spectrum[25]. Sr cations at the $A$-site are observed to preserve their oxidation state, consistent with the Bader charges analysis. The XPS spectrum of Co 2p is shown in Figure 7 (b), which shows the co-existence of $Co^{3+}$ and $Co^{2+}$. The deconvoluted XPS peaks at binding energies 779.56 and 795.28 eV ascribed to the $2p_{3/2}$ and $2p_{1/2}$ electrons, respectively, of trivalent $Co^{3+}$. The peaks at 781.41 and 796.66 eV are attributed to the $2p_{3/2}$ and $2p_{1/2}$ electrons, respectively, of the bivalent $Co^{2+}$. The peaks appearing at 786.38 and 801.05 eV are attributed to 80 and 40 nm thin films[76–78], respectively. The concentration of $Co^{3+}$ is observed to be higher than the $Co^{2+}$ state in thin films as compared to bulk powder. This occurs to maintain the charge neutrality in the structure[79]. The XPS analysis shows that the change in valence happens exclusively at the Co-site. We observed two peaks for Nb at 206 and 208.8 eV associated with $3d_{5/2}$ and $3d_{3/2}$, respectively (Figure 7 (c)). The symmetric nature of Nb XPS predicts +5 single oxidation state[80,81]. In line with the bader charge analysis, Co cation is observed to be in a multivalent oxidation state, showing its redox ability, whereas Nb is observed with only +5 oxidation state, indicating its redox inability. The O 1$s$ spectra of the as-synthesized SCNO indicates the presence of oxygen vacancies (Figure 7 (d)). The O 1$s$ core-level spectra of oxygen lattice ($O_{lattice}$) and adsorbed oxygen ($O_{adsorbed}$) are found to be at 528.70 and 531.21 eV, respectively for SCNO thin films[82–85].

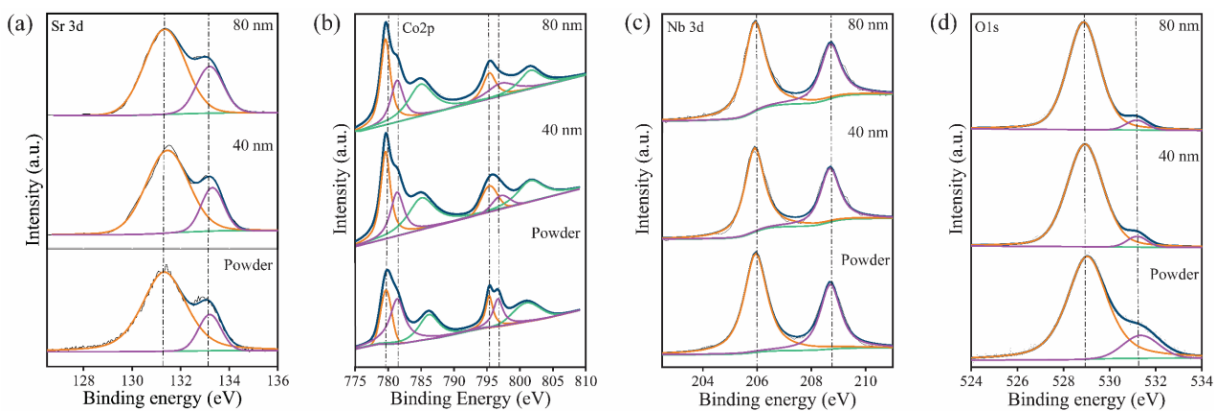

**Figure 7.** XPS spectrum for (a) Sr, (b) Co, (c) Nb, and (d) O in SCNO powder, and 40 and 80 nm thin films.

The ORR activity of SCNO was investigated using the symmetric cell configuration of SCNO||SDC||SCNO. The EIS of symmetric cells contains information about processes involved in ORR, which are further analyzed by using the DRT technique[55]. The DRT deconvolutes the EIS response by separating the ORR processes into a number of sub-steps like gas diffusion, $O_2$ surface adsorption, adsorption-dissociation, charge transfer, and oxygen ion migration from the electrode to the electrolyte interface. The semi-arc in EIS represents the polarization resistance. DRT response indicates the peaks determining the rate-controlling steps involved in the polarization response. The EIS of a symmetric cell with thin film electrodes of 40 nm and 80 nm thickness and a porous electrode are shown in Figure 8 (a-c). The polarization resistance ($R_p$) for 40 and 80 nm dense thin film symmetric cells are observed between 0.329 - 0.241 $\Omega$ cm$^2$ and 1.095 - 0.438 $\Omega$ cm$^2$, respectively, in the temperature range of 773 - 973 K in air atmosphere. The results indicate higher ORR activity with increasing temperature and decreasing thin film thickness. There is an absence of the Warburg-type component in polarization spectra, which indicates that the electrode's dense thin film microstructure is unlikely to promote gas-phase migration to the triple-phase boundary (TPB).

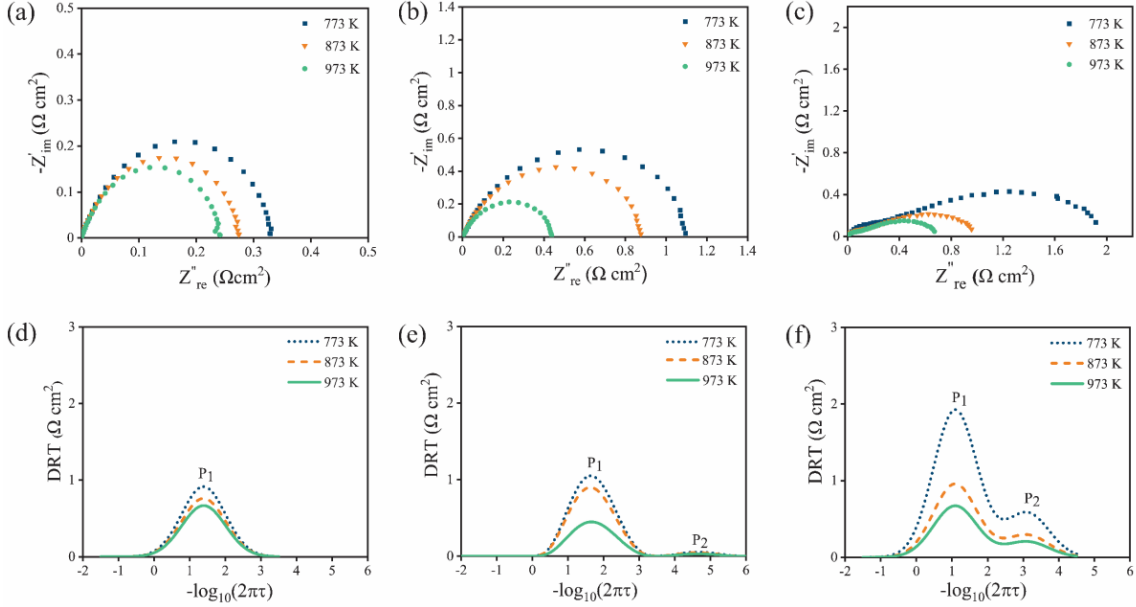

**Figure 8.** Impedance spectra and DRT analysis of (a & d) ~40 nm (b & e) ~80 nm dense thin film, and (c & f) porous electrode of SCNO double perovskite symmetric cell.

Furthermore, the impedance spectra shown in Figure 8 (a, b) were deconvoluted using the DRT method. The DRT technique identifies and separates individual sub-processes by their distinct relaxation timings[86,87]. The low-frequency P1 is associated with oxygen adsorption, bulk diffusion, and catalytic dissociation. The intermediate frequency P2 is correlated with the charge transfer, whereas the high-frequency P3 is associated with the interfaces between electrodes and electrolytes[88,89]. Figure 8 (d, e) illustrates the DRT curve for dense thin film microstructure electrodes with 40 and 80 nm thicknesses, respectively. The single peak P1 was observed in the DRT response of a 40 nm thin film symmetric cell, which is associated with oxygen ion transport that predominantly takes place through the bulk diffusion channel[90,91]. When the temperature increases, there is a noticeable reduction in the peak area, indicating an enhancement in the activity of the ORR. In 80 nm thin film DRT, a peak P2 was observed in a very small magnitude and demonstrated a wide association with higher frequency phenomena, such as the transport of ions and the transfer of charges at the interfaces between the electrode and electrolyte. Qiu *et al.* observed a comparable pattern indicating that an increase in electrode thickness resulted in the emergence of a rate-limiting process for the ORR[92]. According to the aforementioned hypothesis, the dissociation of oxygen species and their subsequent migration through oxygen vacancies occurred due to bulk diffusion through the thin film electrode toward the electrolyte[93].

Furthermore, Figure 8 (f) displays the DRT curve of a porous electrode symmetric cell that encompasses two distinct processes, P1 and P2. The process involves the adsorption and diffusion of oxygen molecules across the electrode surface, where they subsequently break into oxygen atoms. The dissociation takes place followed by bulk diffusion and charge transfer mechanism, which is followed by the reduction of oxygen ions at the porous electrode's TPB. The DRT curve demonstrates that the low-frequency processes, specifically P1, exert a more substantial impact on the overall electrochemical reaction compared to the high-frequency process, P2. Therefore, the rate-limiting mechanism pertains to the absorption of oxygen molecules, their diffusion across the electrode surface, and subsequent dissociation. The findings on dense thin film electrodes indicate that the rate-limiting step is associated with only bulk diffusion, and the porous electrode contains TPB that shows the occurrence of multiple processes for oxygen ion transfer[94].

As SCNO is found to have a high surface stability, ORR activity of SCNO is studied in full cell configuration. A full cell with a NiO anode support was constructed to assess the SOFC performance of the cell. The cell was built by applying a GDC electrolyte layer with a thickness of approximately 500 μm, followed by a SCNO cathode. The cell was then tested at temperatures ranging from 773 – 973 K. The anode was exposed to a 3% humidified $H_2$ fuel, while the cathode was exposed to air. The full cell open circuit voltage (OCV) is observed at roughly 0.981 V at 973 K. This observation indicates that the cell is effectively sealed. Figure 9 (a) shows the full cell's current density-power density (I-P) and current density-voltage (I-V) curves. The SCNO cathode exhibits peak power densities of 0.421, 0.540, and 0.633 W/cm$^2$ at 773, 873, and 973 K, respectively. The full cell performance of SCNO is observed to be enhanced as compared to other double perovskite oxides probed in literature (Table S6). The EIS was conducted under the OCV condition, Figure 9 (b). The $R_p$ extracted from the Nyquist plot were observed as 0.509, 0.407, and 0.254 Ω cm$^2$ at 773, 873, and 973 K, respectively. The observed enhancement in electrochemical performance at intermediate temperatures can be attributed to the increased diffusion of ions. We observed a decrease in the resistance encountered during charge transfer and a corresponding enhancement in the activity of the ORR.

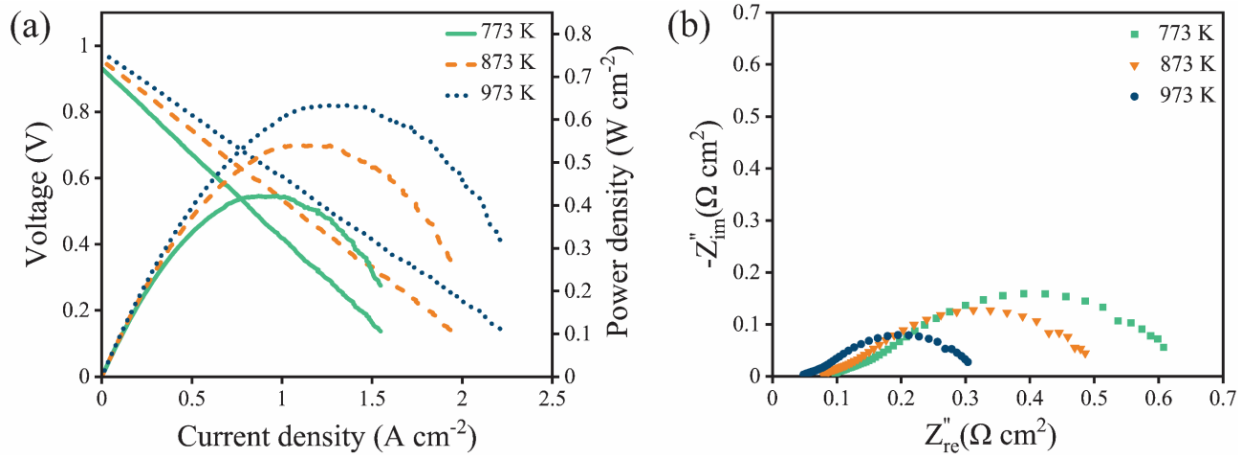

**Figure 9.** (a) I-P, I-V curves, and (b) EIS of anode-supported NiO-GDC|GDC|GDC-SCNO.

## 4 Conclusions

SCNO perovskite oxide is investigated as a promising cathode material for IT-SOFCs. Despite the presence of Sr at the A-site, SCNO is observed to demonstrate a high surface stability under SOFC operating conditions. A combination of DFT and MD simulations show that Nb at the B′-site, effectively suppresses Sr-cation segregation, enhancing material stability. Our structural studies, simulating the impact of GDC substrate compression using DFT and MD, further indicate a reduced Sr-cation segregation. However, Nb's low electronegativity, stable configuration, and less reducible nature also hinder the formation of oxygen vacancy. On the other hand, the highly redox-active Co at the B-site is found to favor oxygen vacancy formation. Hence, Co at the B-site and Nb at the B′-site together position SCNO as a strong candidate with enhanced surface stability and electrocatalytic performance.

Further, SCNO was synthesized via an auto-combustion method, and characterization using X-ray diffraction confirmed successful phase formation in both the SCNO powder and thin films deposited on GDC. The observed oxidation states of elements in SCNO using XPS were found to be consistent with the DFT-based Bader charge analysis. Electrochemical testing of a single cell with SCNO as the cathode yielded a power density of 0.633 W/cm² at 973 K. The ORR mechanism was analyzed using EIS and DRT analysis. We observed bulk diffusion as the primary performance-limiting factor, especially with increasing electrode thickness. This study provides insights into identifying performance-limiting mechanisms through a combined theoretical and experimental approach and positions SCNO as a high-performance cathode material for IT-SOFCs.

## Conflicts of interest

The authors declare no conflicts of interest. They have no known financial interests or personal relationships that could have influenced the work presented in this paper.

## Data availability

The data supporting this article has been included as part of the Supplementary Information.

## Acknowledgments

JK acknowledges the financial support from the Council of Scientific & Industrial Research (CSIR), India [Grant No. 09/086(1412)/2019-EMR-I]. MAH acknowledges the support from the Department of Science and Technology, India, under DST/TDT/AM/2022/138 (G)/2. The results presented in this paper are derived from computations performed on the High-Performance Computing cluster, Padum, at the Indian Institute of Technology Delhi and PARAM Rudra supercomputing facility at Inter University Accelerator Centre Delhi.